\documentclass{elsart3p}   %3p
\usepackage{graphicx,amssymb}

\usepackage[T1]{fontenc}
\usepackage[english]{babel}

\def\a{\alpha}          \def\b{\beta}            \def\g{\gamma}
          \def\eps{{\epsilon}}     \def\ve{\varepsilon}
         \def\k{\kappa}           
                           
\def\r{\rho}                       
\def\vphi{\varphi}      
          \def\D{\Delta}

\def\be{\begin{equation}}            \def\ee{\end{equation}}
\def\ba#1{\begin{array}{#1}}         \def\ea{\end{array}}

\def\vv{\vartheta}

\journal{}
\begin{document}
%\runauthor{aaa}
\begin{frontmatter}
\title{Anisomagnetic quasi-achromats with
small effective emittance}
\author{I.L.\,Zhogin\thanksref{cor}}
\ead{zhogin@mail.ru}
\address{Institute of Solid State Chemistry \& % and
Mechanochemistry, SB RAS,
  Kutateladze str. 18, 630128
 Novosibirsk, Russia }
 %Federation
\thanks[cor]{Corresponding author. Tel.: +7 383 3294298}

\begin{abstract}
Quasi-achromat lattices (small dispersion is allowed in their
straight sections, between their cells) are considered; in a cell,
there are bending magnets of two kinds, of unequal magnetic field.
Minimization of the effective emittance
 is carried out by the
following algorithm (which follows Teng, and partly Lee).

(1) Every inner dipole's contribution to the natural emittance is
minimized with respect to all optics parameters (relating to
dispersion and beta-function), except for the shift parameter,
$s_0$, which specifies the interval between the beta-function
minimum and the center of a magnet; for a side bending magnet, its
contribution to an integral {\em relating} to the effective
emittance (the {\em relation} uses the fact that the {\em
arithmetic mean}
 majorizes the {\em geometric mean}) is minimized.
 (2) The other parameters of dipoles (fields, lengths
 or angles ratios, shifts) are
restricted with the boundary conditions: the equality of
Courant-Snyder invariants on the exit from a magnet and on the
entrance to the following one. (3) The minimum of effective
emittance and the last free parameters (two or three) can be found
by computation.

The accuracy of this method falls with decreasing of the number of
internal dipoles in a cell, and still the isomagnetic Tanaka-Ando
minimum  (for the modified DBA$\!{}^{*}$-lattice which has no
inner dipoles) is reproduced with accuracy better than half a
percent. If the number of dipoles per cell does not exceed four,
the smallest effective emittance (14\% lower than TA-limit) is
achieved for QBA$\!{}^{**}$-lattice where all dipoles have nonzero
shifts.

\end{abstract}
\begin{keyword}
effective emittance reduction \sep storage ring  \sep
%%detonation front;
anisomagnetic lattice \sep Courant-Snyder invariant
\PACS
\end{keyword}
\end{frontmatter}

\section{Introduction}
Growing requests and needs of users of synchrotron radiation
   call for further improvements of parameters of specialised
 synchrotron light sources. One of the most important parameters
  is the
 horizontal natural emittance,  $\eps_x$ (it also defines,
 through some coupling, the vertical emittance).
 Its value depends on the magnetic structure (lattice) of a
 storage ring, and first of all -- on parameters and arrangement
 of dipole magnets, which should provide a closed orbit for the
 electron beam.
 For achromatic lattices
  DBA (double bend achromat -- two dipole magnets per cell),
  TBA, QBA, where the `intercell' regions have zero dispersion,
 the conditions of  natural emittance minimization are well known
 from the works  of Teng, and Lee, \cite{teng,lee}.
 The emittance strongly decreases with increase of the number of
 dipole magnets in a ring,
 \[ \eps_x \propto N^{-3} \, ;\]
  at that, the size and costs of storage ring grow with $N$ as
 well.

 The figure of merit of an insertion device (ID), serving as a SR-source,
 is defined, however, by the {\em effective emittance}, which depends on
 dispersion
 (or Courant-Snider invariant) in the ID straight section.
 It turns out that the effective emittance can be lowered
   if to weaken
 the dispersion-free condition  \cite{rop,tan}. Therefore,
 the beam optics of some electron rings (Elettra, ESRF), initially
 achromatic, of  DBA kind, was modified
 (so called distributed dispersion lattice, or quasi-achromat, DBA*).

 With the aim to further reduce the emittance in ESRF, together
 with the doubling of the number of dipoles, from 64 to 128, they
 are considering the possibility to use dipoles with
  varying (along the orbit)
 magnetic field   \cite{rop,papa}.

In this work the problem of minimization of the effective
emittance is to be solved for the case when the lattice cell is
symmetric (with respect to inversion in its center) and includes
magnets of two types: {\em internal} dipoles $M_0$ and a pair of
 the {\em end\/} (or side) dipoles
   $M_1$; the total number of magnets in a cell is
    $m=m_0+2$.
 Modified lattices  TBA* and DBA* correspond to
    $m_0=1$ and $m_0=0$, respectively. In the case $m_0=2$, the
    internal dipoles can also have non-zero shift; this case will
    be denoted as QBA**.

 Dipoles with unequal fields serve as SR sources
 with  different critical energy, i.e, with peak regions in
 different parts of X-ray spectrum, so they can serve better
 for different  `classes' of SR users.
 However, large number of free parameters, which describe
 anisomagnetic cell, makes the search of the best parameters (when
 the effective emittance is minimal) a very difficult problem.

 Here a possible way, an algorithm to solve this problem is
 suggested, which divides the problem into a few more simple parts.
  In principle, this approach allows to consider more complex
  lattice, with three or even four kinds of dipole magnets.

\section{Effective  emittance}
 The horizontal emittance of electrons,
 $\eps_x$, and the effective (horizontal)
 emittance,  $\eps_{\rm eff}$, are defined by the next
 expressions, relating to some integrals along the orbit
   (they depend on the orbit's radius,
  $\r$, and optics function, including $\b$-function)
 \cite{lee,rop,tan}:
\be \label{emi}
\eps_x=C_q\,\g^2\, \ve_x , \
 \ve_x=\frac{I_5}{J_xI_2}=\frac{I_5}{I_2-I_4},
 \ee
\be\label{emi-eff} \eps_{\rm eff}=C_q\,\g^2\, \ve_{\rm eff}\, , \
\ve^2_{\rm eff}=\ve_x\left(\ve_x+ \frac{H_{\rm
ID}\,I_3}{2I_2+I_4}\right). \ee
 Here $\g$ is relativistic factor,
 \[C_q=3.84\times 10^{-13}\,\mbox{m}, \ \,
C_q\g^2= 1470\,{\rm nm}\,
 (E[{\rm GeV}])^2 ;\]
 these are the usual notations of some ring integrals:
 \be \label{II}
\begin{array}{l}
  I_2  =\oint \r^{-2}\,ds, \ \
I_4 =\oint \eta \r^{-3}(1+2k \r^2) \,ds, \\
 I_3 =\oint |\r|^{-3}\,ds\, , \
I_5=\oint |\r|^{-3}H \,ds \, ;
\end{array} \ee
$H_{\rm ID}$ in (\ref{emi-eff}) is the Courant--Snyder invariant
in the region of insertion devise,
 \be\label{ics}
 H= \b^{-1}\{\eta^2+(\a
\eta+\b \eta')^2\}, \ee
 $ \a=-\frac12 \b'$; $\eta$ is dispersion function.

 Some rings, sources of SR, were designed that the regions of ID
 would have invariant   $H_{\rm ID}$ vanishing
 (i.e.,  dispersion and its derivative are both zero), but in the
 curse of time, lattice parameters were changed to reach better
 minimization  of the effective emittance (\ref{emi-eff}).

Integral $I_4$ is small in comparison with  $I_2$, so it can be
 neglected,  in the case of separate function magnets
  (dipoles have no quadrupole  component, and $k=0$).

 So, we seek to minimize the next combination
 of ring integral:
 \be\label{emi-eff2}
\ve^2_{\rm eff}= \frac{I_5I_6}{I_2^2}=\frac{I_5(I_5 + H_{\rm
ID}\,I_3/2)}{I_2^2}\, ; %%
\ee
 and here one should note that
 integrals $I_2$ and $I_3$ depends only on the field of dipoles
  (radius of the orbit), but does not depend on betatron and
  dispersion functions.
  On the other hand, integrals  $I_5$ and $I_6=I_5+H_{\rm
ID}\,I_3/2$ are a sum of positive contributions from all magnets
 (because the Courant-Snyder invariant is always positive).

 Therefore this problem of optimization
 can be attempted to be divided into a few relatively simple
 steps, that is, to be factorized:

(1) Firstly, we make minimization of a magnet's contribution to
 $I_5$ with respect to parameters
$\b_0,\, \eta_0,\, \eta'_0$, i.e., the value of beta-function at
its minimum, and the dispersion functions in the point of this
minimum (this point serves as the reference point, zero point
 of the orbit coordinate $s$  in a dipole); for the end magnets,
 which neighbour ID straight sections and directly define invariant
  $H_{\rm ID}$, we minimize their contribution to the sum
$I_5+I_6$, %%=2I_5+H_{\rm ID}\,I_3/2$,
because it \emph{approximates\/} in a good way the necessary
minimization of the product $I_5I_6$ (the arithmetic mean
 majorizes the  geometric mean).

(2) The other dipole's parameters -- the curvature $\r^{-1}$ (or
the field), the length or the angle of rotation, $l=\vv \r$, as
well as the the coordinate of dipole's center, $s_0$, or the
(dimensionless parameter) \emph{shift}, $x=-2s_0/l$, --
 should be restricted with the matching condition, which states
 the equality of the Courant--Snyder invariants on the exit from a
 dipole and on the entrance to the next one.
 One more restriction is that the orbit is closed, i.e.,
 the sum of angles
 $\vv_i$ of all dipole magnets is equal to $2\pi$.

(3) After eliminating the angles, one can minimize the effective
emittance with respect to the rest (two or three) parameters:
\[ \k=\r_1/\r_0,\
%% \k_2=\r_0/\r_2,\
x=x_0, \ y=x_1 \] (the internal magnets can have nonzero shift
 $x_0$, if their number is two).

It is well known that the contribution of a dipole magnet to the
(natural) emittance is minimal when its center coincide with the
minimum of beta-function  (besides, this minimum and the
dispersion parameters should have definite values depending on
the magnet's field \cite{teng,lee}). Such a magnet can be called
{\em symmetric\/}  (of symmetric arrangement), or {\em unshifted};
 otherwise, we have a magnet with a shift.

As a rule, lattices consist of repeating, equal collections of
dipole magnets -- cells or periods (or superperiods).
 We will consider quite general case of a periodic lattice of
  $n$ symmetric cells  (inversion in the cell center).
 Every cell contains $m$
dipole magnets, including $m_0=m-2$ internal magnets $M_0$
(unshifted, as a  rule),
 and two (one on each side) end magnets
 $M_1$, with a shift  $x_1$.
Considering the right half of the cell, we will  imply that the
shift of the end magnet is positive  (and it is negative in the
left half of the cell). The total number of dipoles is $N=n\cdot
m$; their rotation angles, $\vv_0,\,\vv_1$,
 should comply with the requirement:
\be\label{vvs} \sum \vv_i=n(m_0\vv_0+2\vv_1)=2\pi\,. \ee
 Let us introduce the notation of the {\em mean} angle:
  $\bar{\vv}=2\pi/N$.

\section{Minimization of contribution (to the emittance)
of  the internal dipoles}
 Let the length of a magnet is $l=\r \vv$, and the coordinate of
 its center (from the minimum of beta-function) is $s_0=-xl/2$.

Following  Teng \cite{teng}, one can write the contribution of a
magnet to  integral  $I_5$, as well as the Courant--Snyder
invariant at its right and left edges,  $H_{\pm}$, in the next
form  (see the expression (9)--(11) from \cite{teng}):
\be\label{hpm} H_{\!\pm}=\b_0(\eta'_0+\sin \vphi_{\!\pm})^2
+\frac{\r^2}{\b_0}\left(\! 1-\frac{\eta_0}{\r}-\cos
\vphi_{\!\pm}\!\right)^{\!2}\!\!, \ee \vspace{-1mm}
\begin{eqnarray}
\D I_5 &\equiv &\frac1{\r^3}
 \int_{s_-}^{s_+}\! H\,ds
= \frac{\vv\b_0}{\r^2}[A+D+
(\eta'_0+E\sin \vphi_0)^2]\nonumber\\
&+& \frac{\vv}{\b_0} [A-D+(1-\frac{\eta_0}{\r}
-E\cos\vphi_0)^2], \label{di5}
\end{eqnarray}
\begin{eqnarray}
&&\mbox{where }
 s_{\pm}=s_0\pm \frac l2, \
 \vphi_{\pm}=\frac{s_{\pm}}{\r} , \
 \vphi_0=\frac{s_0}{\r}=-\frac{x\vv}{2},
  \nonumber \\
  &&\!\left\{\!\!
\begin{array}{l}
 A=\frac12-\frac{1-\cos \vv}{\vv^2}
\approx \frac{\vv^2}{4!}-\frac{\vv^4}{6!} ,\\[5pt]
B=\frac12-A-\frac{\sin\vv}{2\vv}
\approx \frac{\vv^2}{4!}
-\frac{2\vv^4}{6!}  ,
\\[5pt] D=B\cos 2\vphi_0\approx
\frac{\vv^2}{24}\left[
1-\frac{\vv^2}{30}(2+15x^2)\right]
,\\[5pt]
 E=\frac{\sin(\vv/2)}{\vv/2}
 \approx 1-\frac{\vv^2}{24}
 +\frac{\vv^4}{16\cdot5!} .
\end{array}
\right. \label{ADE}
\end{eqnarray}
The last term in the last equation is added to show how quickly
decreases the contribution of the further series terms.
 It is assumed that angles $\vv_i$ are small enough (for all dipoles
 of the storage ring), so already the first
 terms of angle expansion gives a sufficient accuracy.

Minimization of expression (\ref{di5}) with respect to $\eta_0, \,
\eta'_0$ is reached if these parameters complies with the next
conditions \cite{teng}:
 \be\label{etas} \left\{
\begin{array}{l}
\eta'_0=-E\sin\vphi_0\approx
\frac{\,x\vv}2
%%\left(1-\frac{1+x^2}{24}\vv^2\right)
,\\[5pt]
\eta_0=\r-\r E\cos\vphi_0\approx
\r\vv^2\frac{1+3x^2}{24} \, .
\end{array}
\right.
\ee

 The dipole's contribution to the integral
 takes the form
\[\D I_5=\frac{\vv}{\r} \left\{
\frac{\b_0}{\r}(A+D)+\frac{\r}{\b_0}(A-D)\right\}\, ; \]
 further minimization with respect to
  $\b_0$ gives now
\begin{eqnarray}\label{beta0}
  \b_0^2 &=& \r^2\frac{A-D}{A+D}\approx
  \r^2\vv^2\frac{1+15x^2}{60}, \\
\label{i5min}
 \D I^{\rm(min)}_{5} &=& \frac{2\vv}{\r}\sqrt{A^2-D^2}
\approx
\frac{\vv^4\sqrt{1+15x^2}}{12\sqrt{15}\r}.
\end{eqnarray}

 Taking into account (\ref{etas}), (\ref{beta0}),
one can reduce equation (\ref{hpm}) to the following one:
 \be\label{hpmm}
H_{\pm} \approx \r\vv^3 \frac{4\mp 15x
+45x^2}{12\sqrt{15}\sqrt{1+15x^2}}\,. \ee

At $x=0$, we obtain the expressions for the central (unshifted)
magnet (the case A of Teng). A lattice composed of  $N$ such
magnets (let $\r=1$, then $I_2=I_3=N\vv, \ I_5=N \D I_5$),
corresponds to the case of minimal natural emittance,
 \be\label{ve0} \ve_0=\frac{\bar{\vv}^{\,3}}{12\sqrt{15}}
\,\left(=\frac{I_5}{I_2}\right); \ \ \bar{\vv}=\frac{2\pi}{N}
\,(=\vv); \ee
 the effective emittance of this lattice
 is equal to $3\ve_0$.

 It is convenient to measure  emittances in units of $\ve_0$,
 through introduction of dimensionless values $f, \,g$.
It is known \cite{lee}, that for the DBA lattice
\[f\equiv \frac{\ve_x}{\ve_0 }=3, \
\ g\equiv\frac{\ve_{\rm eff}}{\ve_0 } =3.\]
 Dipoles of   DBA-lattice have the shift
 $x=\frac14$; using this value $x=\frac14$ in
(\ref{etas})--(\ref{hpmm}), one can obtain (the effective
emittance of such a quasi-achromat lattice is smaller in
comparison with DBA case) $f= \frac{\sqrt{31}}4, \
g=\frac{111}{8\sqrt{31}}\approx2.49.$  At the point of minimum of
$H_+(x)$, see (\ref{hpmm}), $x=0.229$ (this is the root of
equation $45x^3+2x=1$), the effective emittance becomes yet some
smaller: $g=2.43$. %%
 However, the more exact approach to minimize DBA* is to use
 another way to choose the optics parameters -- as for the end
 dipoles.
\section{Minimization of the end dipoles' contribution to the effective
emittance}
 Let the integral over all cell dipoles but the end ones
 is equal to $I^*_5$, and the end dipole's contribution is $\D I_5$.
 Accounting for equation (\ref{emi-eff2}),  we have
\begin{eqnarray*}
 \D (I_5 I_6) &=&n^2(I^*_5+2\D I_5)(I^*_5+2\D I_6)
 -n^2(I^*_5)^2 \\
   &\approx & 2n^2I^*_5(\D I_5+\D I_6)
   \propto 2\D I_5+J H_+/2\,.
\end{eqnarray*}
Here $J=I_3/(2n)$ -- integral along the half-cell, $n$ is the
number of cells. Using (\ref{hpm})--(\ref{ADE}), one can find the
expression, which should be minimized (with respect to dipole's
parameters $\eta_0, \eta'_0, \b_0$; for simplicity sake,
take for a while $\r=1$):
\begin{eqnarray}
\D I_5 &+&\D I_6 = 2\vv\b_0[A+D+
(\eta'_0+E\sin \vphi_0)^2]\nonumber\\
&&+ \frac{2\vv}{\b_0} [A-D+(1-\eta_0
-E\cos\vphi_0)^2]  \label{di56} \\
&+& \frac{J\b_0}{2}(\eta'_0+\sin \vphi_{\!+})^2
+\frac{J}{2\b_0}(1-\eta_0-\cos \vphi_{\!+})^{2}
.\nonumber
\end{eqnarray}
Zeroing the derivatives by $\eta'_0$ and $\eta_0$, one can find
the values of these parameters when the minimum occurs (restore
$\r$):
\be\label{etas2} \left\{
\begin{array}{l}
\eta'_0\approx
\frac{\,\vv}2 \cdot
\frac{4y\vv-J\r^2(1-y)}{4\vv+J\r^2}
\,,\\[5pt]
\eta_0\approx
\frac{\r\vv^2}{24}\cdot
\frac{4\vv(1+3y^2)+3J\r^2(1-y)^2}
{4\vv+J\r^2} \, ;
\end{array}
\right.
\ee
 here $y$ is the shift of end dipoles. Substituting this values to
  (\ref{di56}) one can obtain:
\begin{eqnarray}\nonumber
\D I_5 &+&\D I_6 = \frac{\vv^3}6 \left\{
4\b_0\frac{\vv+J}{4\vv+J} \right. \\
  &+&\left.\frac{\vv^2}{60\b_0}\left[
1+15y^2+\frac{5J(1-3y)^2}{4\vv+J}\right]\right\}
\,;\label{5-6}
\end{eqnarray}
 the minimum is reached if (restore $\r$)
\be \label{beta02} \b^2_0=\r^2\vv^2\frac{2\vv(1{+}15y^2)
+3J\r^2(1{-}5y{+}10y^2)} {120(\vv+J\r^2)}\, . \ee
 The edge CS-invariant takes the form:
 \begin{eqnarray}
% \nonumber
  H_{\pm} &=& \frac{\b_0\vv^2}{4}
  \left[\frac{4\vv+J\mp J}{4\vv+J}\right]^2
  \nonumber\\ \label{hpm2}
  &+& \frac{\vv^4}{9\b_0}
  \left[\frac{\vv\mp 3y\vv+(3\mp 3)y J/4}{4\vv+J}\right]^2\,.
\end{eqnarray}

In principle, now all is ready for the final minimization:
in addition to equations (\ref{5-6})--(\ref{hpm2}) one has to take into
account that
$\D I_6 -\D I_5 = JH_+/2\,.$

\section{Minimization for  DBA*-lattice}
 In order to check the appropriateness of the equations of the
 previous section, let us consider the simplest quasi-achromat lattice,
 with two magnets in a cell.
  In this case (in this section again $\r=1$)
\[ J\equiv I_3/N=\vv=\bar{\vv} ,\]
 and equations (\ref{etas2})--(\ref{hpm2})
 lead to
\[
\left\{
\begin{array}{l}
\eta'_0=\vv
\frac{5y-1}{10}
\,, \ \,
%%\\[5pt]
\eta_0=\vv^2\frac{7-6y+15y^2}{120} \, ,\\[5pt]
\b_0=\frac{\vv\sqrt{1-3y+12y^2}}{4\sqrt{3}}\,.
\end{array}
\right.
\]
\vspace{-10pt}
\begin{eqnarray*}
 \D I_5 &+&\D I_6 =2\vv^4
 \frac{\sqrt{1-3y+12y^2}}{15\sqrt{3}}, \\
  H_+ &=&\vv^3
 \frac{7-33y+72y^2}{75\sqrt{3}\sqrt{1-3y+12y^2}}.
\end{eqnarray*}

As a result, we find the expressions for the `simple'
dimensionless emittances
(natural and effective, respectively), see
(\ref{ve0}):
\be\label{fg0}
\begin{array}{l}
f=\frac{\D I_5}{\vv\ve_0}=
\frac{13-27y+168y^2}{5\sqrt{5{-}15y{+}60y^2}}
\,, \ g^2{=}\frac{\D I_5 \D I_6}{\vv^2\ve^2_0},
\\[7pt]
 g^2=
\frac{3(13-27y+168y^2)(9-31y+104y^2)}
{125(1-3y+12y^2)}\,.
\end{array}
\ee
 \vspace{-4mm}
\begin{figure}[hbt]
\begin{center}   %% 0 0 213 139
\includegraphics*[width=71mm]{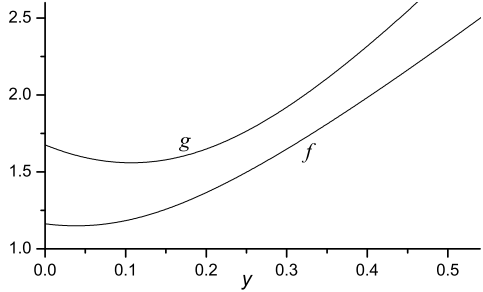}\\[-5mm]
\caption{`Simple' emittances of
DBA* quasi-achromat lattice: $f$ -- natural, $g$ -- effective.}
\end{center}
 \label{fi1}
\end{figure}
The corresponding curves are shown at the fig.~1.
The minimum of effective emittance if reached
 at
$y=0.107$, and its value is $g_{\rm min}=1.559$.
So, the suggested minimization algorithm reproduces with
an accuracy better than half a percent the minimum of
Tanaka--Ando  \cite{tan}, which is 1.552.

It the next section we will consider more complex lattices,
when the cell includes internal dipoles, with zero shift.

\section{Lattices TBA*, QBA*, et cet.}
Let $m_0$ is the number of internal dipoles, and
$m=m_0+2$ is the total  number of dipoles in a cell.
We introduce the next dimensionless parameters:
 \[q=\vv_1/\bar{\vv}, \ p=\vv_0/\vv_1, \
\k=\r_1/\r_0,\]
 and let $\r_1=1$ (unit length).
We will also omit in equations the mean angle
(i.e., taking formally $\bar{\vv}=1$), because it should not enter
into the simple emittance,
both $f$ and $g$.

Still it is impossible to find
  $g(y,\k;\, m_0)$ in an explicit form, but one can find
  this function through a numerical computing, according
  the following algorithm.

The orbit's closure condition and the first integrals
along the half-cell, normalized on $\vv_1$, see (\ref{II}), read:
\[
q=\frac{m}{m_0 p+2}\,, \ \
i_2\equiv I_2/(2n\vv_1)= 1+\frac{m_0}2 p\k\,, \]
\be\label{j}
j\equiv J/\vv_1=I_3/(2n\vv_1)=1+\frac{m_0}2 p\k^2\, .
\ee
Equation (\ref{beta02}) gives (after normalization)
\[b\equiv\frac{\b^{(1)}_0}{\vv_1}=\sqrt{\frac{2+30y^2
+3j(1-5y+10y^2)}
{120(1+j)}}\, .\]
In the same way  (\ref{hpm2})
and (\ref{5-6}) transform to
\[h_{\pm}\equiv\frac{H^{(1)}_{\pm}}{\vv^3_1}, \ \,
S\equiv\frac{\D I^{(1)}_5+\D I^{(1)}_6}{\vv_1^4}=
\frac{4b(1+j)}{3(4+j)}\,.
\]
The matching condition $H_+^{(0)}=H_-^{(1)} $ gives

\be\label{hphm}
\frac{p^3}{3\sqrt{15}\k}=
b\frac{(2+j)^2}{(4+j)^2}+\frac{(2+6y+3jy)^2}
{36b(4+j)^2}\, (=h_-).
\ee
Equation (\ref{hphm})  subject to (\ref{j})
 can be solved numerically; this gives a unique
 solution
  $p(y,\k;\,m_0)$.
 Now, one can find $b(y,\k;\,m_0)$
 (i.e., the minimum of beta-function of the end
 dipole, $\b^{(1)}_0$),
 and, at last, calculate $f$ and $g$:
\be\label{i5-3} i_5=\frac{I_5}{2n\vv_1}=
\frac{q^3}{4}(2m_0\ve_0 p^4\k+2s-jh_+),
 \ee
\be \label{fg3} f=\frac{i_5}{\ve_0i_2}\,, \ \,
g^2=f\!\left(\!f+\frac{q^3j h_+}{2i_2\ve_0}\!\right); \
\ve_0{=}\frac1{12\sqrt{15}}\,. \ee
 Fig.~2 shows the region of the effective emittance minimum,
 $g(y,\k)$, for TBA* lattice.
The  Table contains parameters,
including the natural emittance, for quasi-achromat lattices
  mBA* up to
$m=10$.\footnote{One can find the m-files relating to this
computations on the site zhogin.narod.ru}

\begin{table}
%%  \centering
  \caption{Parameters of quasi-achromats mBA*}\label{t1}
\begin{tabular}{|c|c|c|c|c|c|c|c|c|c|}
  \hline
$m$&2&3&4&5&6&7&8&9&10\\
\hline
$g_{\rm min}$
 &1.559&1.435&1.353&1.297&1.257&1.226&1.202&1.182&1.166\\
$f$ &1.191&1.207&1.196&1.182&1.169&1.156&1.146&1.136&1.128\\
$y$ &0.107&0.157&0.185&0.203&0.215&0.224&0.231&0.236&0.240\\
$\k$& --&0.850&0.835&0.830&0.831&0.832&0.835&0.839&0.843\\
$p$ & -- &1.134&1.170&1.198&1.222&1.241&1.258&1.272&1.285\\
% 2& 1.559&1.191&0.107& --& -- \\
% 3& 1.435&1.207&0.157&0.850&1.134\\
% 2& 1.353&1.196&0.185&0.835&1.17\\
% 3& 1.297&1.182&0.203&0.830&1.198\\
%    4&1.257&1.169&0.215&0.831&1.222\\
%    5&1.226&1.156&0.224&0.832&1.241\\
%    6&1.202&1.146&0.231&0.835&1.258\\
%    7&1.182&1.136&0.236&0.839&1.272\\
%    8&1.166&1.128&0.240&0.843&1.285\\
  \hline
\end{tabular}
\end{table}

 \begin{figure}[ht] %%f4 111 271 501 511
\hspace{-2mm}      %% 87  262 507 578
\includegraphics[width=80mm]{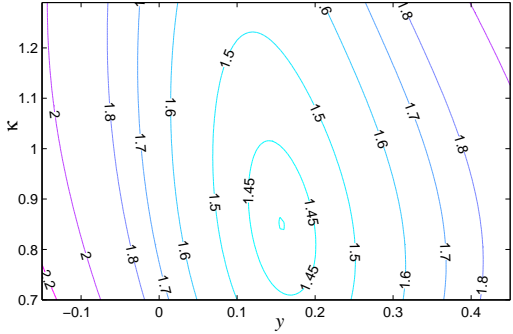}\\[-5mm]
  \caption{ Effective emittance
   $g(\k,y)$ for  TBA$\!{}^{*}\!$ lattice.}
\label{fi2}
\end{figure}
%%%%%%%%%%%%%%%
%%%%%%%%%%%%%%%
\section{Lattice QBA**}
If $m=4$, i.e., four dipoles in a cell,
the internal dipoles can also have a non-zero shift,
  $x_0=x$ (the shift of side dipoles is $x_1=y$).
 This time, one should change the left hand side
 of equation (\ref{hphm}) using equation
   (\ref{hpmm}):
\[ \frac{p^3(4- 15x +45x^2)}{12\k\sqrt{15}\sqrt{1+15x^2}}
=h_-(j,y).\] Moreover, in eq-n (\ref{i5-3})  for $i_5$, the
contribution of internal dipoles should be changed
[see(\ref{i5min})]:
\[ i_5=
\frac{q^3}{4}(2m_0\ve_0 p^4\k\sqrt{1+15x^2}+2s-jh_+).
 \]
Now we need to find a minimum of the function of three variables,
 $g(y,x,\k)$. One can consider two-dimensional
 sections, for different values of  $\k$, seeking for
 the minimum
$g_{\rm min}(\k)$; the plot of this function is shown on
the fig.~3.

Fig.~4 shows the region of $g$-minimum for the case $\k=0.74$
(one can compare with the fig.~2). The minimum itself,
 $g_{\rm min} =1.345$,
 is reached at $y=0.181,\ x=0.046$ (at that $p=1.184$).

\vspace{-3mm}
 \begin{figure}[ht]
\begin{center}
\includegraphics[width=70mm]{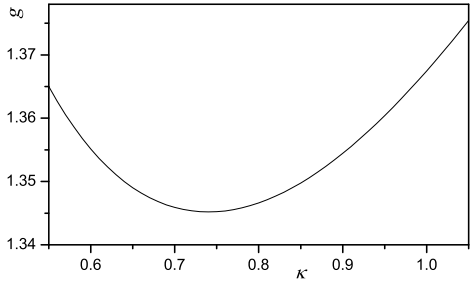}\\[-5mm]
 \caption{Lattice QBA**: emittance $g_{\rm min}$
 vs. $\k=\r_1/\r_0$.}
\end{center}
 \label{fi3}
\end{figure}

 \begin{figure}[ht]
\begin{center}     %87   262   507   578
\includegraphics[width=81mm]{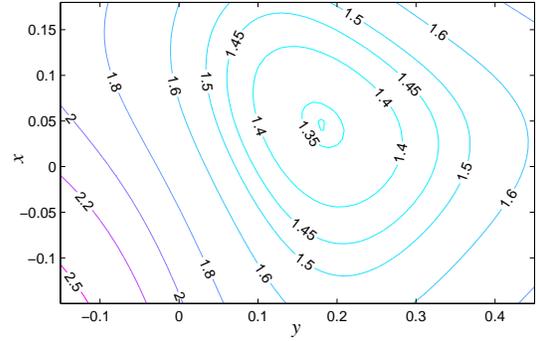}\\[-2mm]
\caption{QBA$\!{}^{**}$:  $g(y,x)$ near its minimum;
 at $\kappa=0.74$.}
\end{center}
 \label{fi4}
\end{figure}

\section{Conclusion}
Anisomagnetic lattices with two sorts of magnets, TBA*, QBA*,
and especially QBA** (internal magnets are shifted too)
can have a bit smaller effective emittance, below the
Tanaka--Ando minimum $g_{\rm TA}=1.552$; in the case of QBA** --
 up to $g_{\rm min}=1.345$.
 It is interesting that the new minimum is reached
 when the field of the internal magnets is smaller,
  at the level 74\%,
  than that of the side magnets
 (so, one can consider the variants
 when internal magnets have quadrupole and/or sextupole
 component).
 One can consider more complex lattices, with tree kinds of
 magnets; note that the matching condition between internal dipoles
 is much simpler than between internal and end magnets.

 The region of minimum is of clear advantages, like the better
 stability with respect to different tolerances, misplacement and
 misalignment errors, and so on,  -- because all gradients are small
 there.

 The suggested minimization algorithm is heuristic in a sense;
 perhaps, it could be a bit complicated.
 For the end magnets, instead of the sum
 $2\D I_5 + H_+J/2$, one can minimize a  general
 linear combination
 $2\D I_5 + \lambda H_+J$; the additional parameter
  $\lambda\approx 0.5$  should be chosen to lower the
  minimum   $g_{\rm min}$.

I express my thanks to K.\,V.~Zolotarev, who has attracted my
attention to this problem, for  interest to this work and a number
of useful remarks.

\end{document}